\documentclass[11pt]{article}
\usepackage{graphicx,amsmath,amsfonts,amssymb}
\usepackage{color,multicol,multirow}
\usepackage{array}
\usepackage{textpos,enumitem,marginnote}
\def\sloppy{\tolerance=100000\hfuzz=\maxdimen\vfuzz=\maxdimen}
\def \beq  {\begin{equation}}
\def \eeq  {\end{equation}}
\def \beqar {\begin{eqnarray}}
\def \eeqar {\end{eqnarray}}
\def\bsp{\beq\begin{split}}
\allowdisplaybreaks
\mathchardef\mhyphen="2D

\def\S {{\cal S}}
\def\la {{\langle}}
\def\ra {{\rangle}}
\def\vx {{\vec x}}
\def\vy {{\vec y}}
\def\vk {{\vec k}}
\def\vf {{\varphi}}

\def\Tr {{\rm Tr}}

\def\vk {\vec{k}}
\def\vx {{\vec x}}

\def\vy{\vec{y}}

\def\del {\partial}

\def\a {\alpha}

\def\E {{\cal E}}

\def\K {{\cal K}}

\def\vf {{\varphi}}

\def\half{\textstyle{1\over 2}}

\begin{document}
\def \CMP {{Commun. Math. Phys.}}
\def \PRL {{Phys. Rev. Lett.}}
\def \PL {{Phys. Lett.}}
\def \NPBProc {{Nucl. Phys. B (Proc. Suppl.)}}
\def \NP {{Nucl. Phys.}}
\def \RMP {{Rev. Mod. Phys.}}
\def \JGP {{J. Geom. Phys.}}
\def \CQG {{Class. Quant. Grav.}}
\def \MPL {{Mod. Phys. Lett.}}
\def \IJMP {{ Int. J. Mod. Phys.}}
\def \JHEP {{JHEP}}
\def \PR {{Phys. Rev.}}
\begin{titlepage}
\null\vspace{-62pt} \pagestyle{empty}
\begin{center}
\rightline{CCNY-HEP-13/6}
\rightline{October 2013}
\vspace{1truein} {\Large\bfseries
Quantum field theories with boundaries and novel instabilities}\\
\vskip .1in
{\Large\bfseries ~}\\
\vskip .1in
{\Large\bfseries ~}\\
{\large\sc T.R. Govindarajan$^a$} and
 {\large\sc V.P. Nair$^b$}\\
\vskip .2in
{\itshape $^a$ Chennai Mathematical Institute\\
Siruseri, TN 603103\\
India}\\
\vskip .1in
{\itshape $^b$Physics Department\\
City College of the CUNY\\
New York, NY 10031}\\
\vskip .1in
\begin{tabular}{r l}
E-mail:&$^a$~{\fontfamily{cmtt}\fontsize{11pt}{15pt}\selectfont trg@cmi.ac.in}\\
&$^b$~{\fontfamily{cmtt}\fontsize{11pt}{15pt}\selectfont vpn@sci.ccny.cuny.edu}
\end{tabular}

\fontfamily{cmr}\fontsize{11pt}{16pt}\selectfont
\vspace{.8in}
\centerline{\large\bf Abstract}
\end{center}
\fontfamily{cmr}\fontsize{11pt}{16pt}\selectfont
Quantum physics on manifolds with boundary brings novel aspects due to boundary conditions.
One important feature is the appearance of localised negative eigenmodes for the Laplacian on the boundary. These can potentially lead to instabilities. We consider quantum field theories on such manifolds and interpret these as leading to the onset of phase transitions.

\end{titlepage}

\pagestyle{plain} \setcounter{page}{2}
\setcounter{footnote}{0}
\setcounter{figure}{0}
\renewcommand\thefootnote{\mbox{\arabic{footnote}}}
\section{Introduction}
In this paper, we consider the problem of general boundary conditions for quantum fields 
defined on a manifold with a boundary. Such manifolds are not only of mathematical interest,
but physically required in several condensed matter systems as well as semiclassical gravity and string theory.
For simplicity, one might even start by
considering a free scalar field $\phi$ with a kinetic term which is given by the Laplacian acting on
$\phi$. The choice of boundary conditions must be consistent with the self-adjointness requirements on the Laplacian and hence are generally described by the von Neumann theory of self-adjoint extensions \cite{vonNeumann}. This theory has recently been elegantly rephrased in \cite{asorey1} and has naturally led to a framework for analyzing the effects of boundary conditions which are more general than Neumann, Dirichlet.
Berry used general Robin boundary conditions to explain novel
behaviour of the spectrum \cite{berry}. One of us used Robin boundary conditions
to obtain novel bound states localised on the boundary to understand blackhole
entropy\cite{trg}.
These boundary conditions have effects on the Casimir energy and 
this has been  exhaustively analysed in \cite {asorey2, KN1}. 
There are a number of related variants which have also been studied before.
Partially transparent boundaries for scalar fields \cite{M-CGM} and for the electromagnetic case
\cite{parashar} have been investigated.
The case when boundary conditions (which can also lead to instabilities as explained below)
can be modeled via $\delta$-functions have also been considered \cite{bordag}.

The set of boundary conditions is given by the choice of a unitary operator $U$, or by the
hermitian operator $\K$ which is its Cayley transform,
on the boundary values of the fields viewed as elements of a Hilbert space of 
$L^2$-functions on the boundary. (We emphasize that one could have more general boundary values for fields which are not square-integrable, singular charge distributions on the boundary
being one class of such examples.We will only consider cases which are $L^2$-functions.)
Specifically, the most general boundary conditions are
given by
\beqar
\phi +i \del_n \phi &=& U \, ( \phi - i \del_n \phi )\nonumber\\
(\phi +i \del_n \phi) (x) &=& \oint_y U(x,y)  \, ( \phi - i \del_n \phi )(y)\label{1}
\eeqar
where $\del_n \phi$ denotes the normal derivative of the field. The alternate way to write this in terms of the Cayley transform $\K$ is
\beq
\del_n \phi = -i \left( {U -1 \over U+1}\right)\, \phi \equiv - {\cal K} \, \phi \label{2}
\eeq
The simplest choices, $\K = 0$ and $\K \rightarrow \infty$, correspond to
Neumann and Dirichlet conditions, respectively. These are special points in the space of boundary conditions. The choice of $\K$ being proportional to the identity operator on
the Hilbert space of boundary values is the
the Robin condition.
One could choose more general ones with different eigenvalues
for $\K$ for different modes on the boundary.
The important point is that there are an infinity of choices for 
$\K$ which leads to negative eigenvalues for Laplacian 
associated with eigenmodes which are 
localised close to the boundary. Such novel states have been exploited 
earlier in several areas, like quantum hall effect, topological insulators
and blackhole physics \cite{REFS}. Clearly such modes will also be important for the Casimir effect
and related issues such as the pair production of particles.
A point worth emphasizing is that these modes of negative eigenvalues can occur infinitesimally close 
(in the space of boundary conditions) to
the ``good choices" like Dirichlet or Neumann.
Generically, all such choices lead to instabilities in  many body physics.
Our experience in physics is that whenever instabilities arise, there is a way out, usually via a phase transition or change of ground state. The classic example is, of course, spontaneous symmetry breaking where a negative (mass)$^2$ term signals the phase transition to
a new stable choice of ground state.
The purpose of the present paper is to ask to what extent a similar scenario can work out for instabilities due to the boundary condition.

In most field-theoretic calculations, normally,
the starting point is to consider the theory at zero temperature.
This means that unless excitations are introduced via external sources, the state of interest is the ground state. If needed, this can then be upgraded to finite temperature with all states contributing, each weighted with the corresponding Boltzmann factor. But in the present case, where the notion of a ground state for the field theory is not clear, our
basic strategy will be to consider the partition function at finite temperature and then ask whether it is possible to lower the temperature to zero. We may view the partition function
as given by the functional integration over the fields in Euclidean spacetime with periodicity along the imaginary-time direction. We will then consider conditions under which
the Euclidean functional integral is well-defined. By considering the limit of this
case where the instabilities will begin to appear, we can get an understanding of
how the transition, if nay, should manifest itself.

In Sec.2, we will briefly consider a couple of examples of how the negative eigenvalues arise.
This is meant primarily to set the framework.
In Sec.3, we will present Bose-Einstein condensation in terms of the Euclidean functional integral
at finite temperature and see that a similar condensation is possible in the case of manifolds with boundary due to the presence of
negative eigenvalues for $-\nabla^2$.
 A key issue here is the existence of a conserved
charge (or particle number) with a corresponding chemical potential.
Since the particle number is fixed, an infinite occupation number for the states of negative
energy is not possible and the theory has a many-body ground state.

In the following section (Sec.4), we consider a real massless scalar field. Since there is
no conserved quantum number in this case, the situation is different.
We show how a Euclidean functional can be defined if we impose a set of restrictions
on the theory.
Effectively, at the level of free particles, there is always a finite temperature, which will play the role that the absolute zero of temperature does in normal theories with no negative eigenvalues
for the Laplacian. There should also be an ``unattainability rule" for this value of
temperature, just as the third law of thermodynamics dictates for normal systems.

Once interactions are introduced, the story can change.
We show in Sec.5 how corrections can be calculated in the theory. The modes with negative eigenvalues
lead to a potential which is repulsive near the boundary and
can alter the eigenstates and eigenvalues.
This gives a way of removing singularities in the Euclidean functional integral.
Finally we end up with a discussion of the results 
and future applications in Sec .6.

\section{Examples of negative eigenvalues}

We will start by considering a couple of examples of how negative 
eigenvalues can arise for $-\nabla^2$. Normally this is expected to be a positive definite 
operator. But with boundary conditions which are motivated  by physical reasoning 
and generic, this character changes. This discussion will help to give a 
concrete form to some of the analysis later.

The first example corresponds to the
space ${\mathbb R}^2$ from which a circular disc of radius 
$R$ has been excised.
We consider the eigenfunctions of the Laplace operator with 
the boundary condition $(\del_r \psi + \kappa \psi)\mid_{boundary} = 0$ 
which is known as Robin boundary condition. 
In other words, we choose $\K$ to be the same for all eigenfunctions
and equal to a parameter $\kappa$. Here $\frac{1}{\kappa}$ has the 
dimensions of length. This is the most general rotation-invariant 
boundary condition. (Some clarification may be useful in this context.
Quite generally, with rotational symmetry, the boundary values 
$\phi +i \partial_n \phi$ may be considered as a linear combination of a 
multiplet of functions corresponding to irreducible representations
of angular momentum of the appropriate dimension. 
(The present example is a bit
too simple from this point of view since the boundary is a circle and all irreducible 
representations are one-dimensional. )
The operator ${\cal K}$ would then have eigenvalues which are degenerate for the members of the multiplet. 
More explicitly in $\del_r \phi = -\K \, \phi$, we can expand $\phi$ in terms of angular momentum eigenfunctions. The derivative, being radial, does not mix these eigenfunctions, showing that
$\K$ is diagonal with the same eigenvalue for a given multiplet; the eigenvalues of $\K$ could
be different for different values of the angular momentum for the multiplets.
The simplest case,namely, when $\K$ is independent of angular momentum is when it is the same for all eigenfunctions. This is what we consider. For more on this matter, but phrased in the framework of
heat kernel expansions, see \cite{avramidi}.)
Physically a parameter such as $\kappa$ can arise due to grainy structure
of the materials in condensed matter systems or from Planck length 
which characterises a fundamental length scale in quantum geometry of
spacetime. 

The eigenvalue equation is written as
\beq
\nabla^2 \, \psi = \lambda^2 \, \psi
\label{b1}
\eeq
where we have introduced a minus sign so that negative eigenvalues correspond to
positive values of $\lambda^2$.
Separation of variables in polar coordinates is straightforward and the eigenfunctions are given by
\beq
\psi_n (r, \theta ) = C\, e^{in\theta} \, K_n (\lambda r)
\label{b2}
\eeq
The required boundary condition becomes
\beq
\kappa \, R = {z \, K_n' (z) \over K_n (z)}, \hskip .3in
z = \lambda\, R
\label{b3}
\eeq
This equation can have solutions for negative values of 
$\kappa$, as discussed in
\cite{trg}. If $z_*$ is a solution of this transcendental 
equation, the corresponding eigenvalue is
$\lambda^2 = (z_*^2 /R^2)$. The largest negative eigenvalue is 
$\propto \kappa^2$. Typically one has a finite number 
of such solutions given by the maximum integer of $\kappa\, R$. These are 
localised close to the boundary. For $\kappa =  -\infty$, corresponding to 
Dirichlet conditions, these are exactly on the boundary and decouples from functions
outside the boundary \cite{balrecent}. 

\begin{figure}[t!]
\begin{center}
\scalebox{.50}{\includegraphics{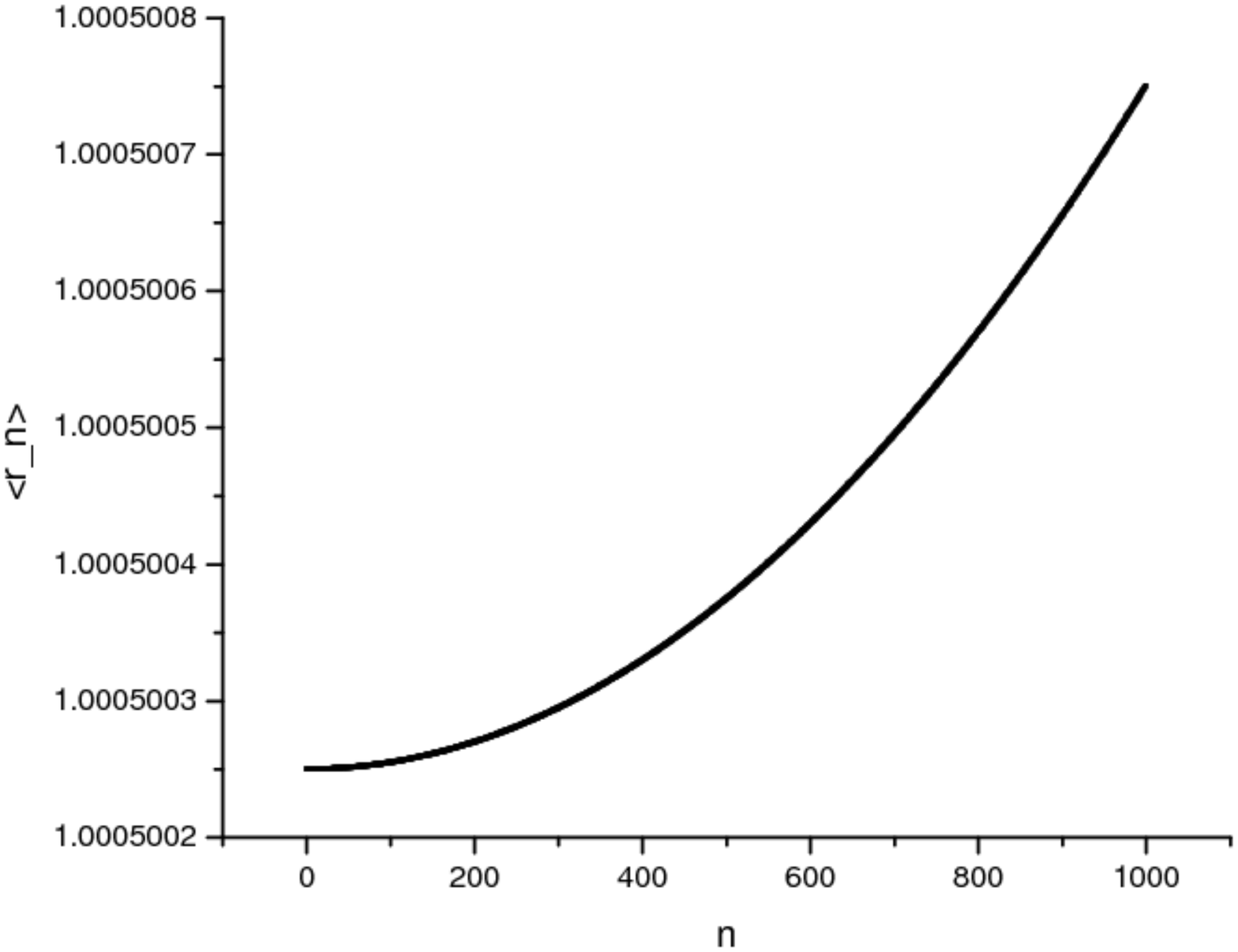}}
\caption{$<r_n>$ vs $n$ for $\kappa = -1000,R~=~1$}
\label{expect}
\end{center}
\end{figure}

In Fig.\,{\ref{expect}} we display $<r_n>, n~=~1,2,...1000$,
the expectation value of $r$ for the $n^{th}$ eigenstate for $\kappa= -1000$ and $R=1$.
It can be seen all the eigenstates are localised within 
$10^{-2}\%$ of the radius and all of them  lie inside shell of thickness 
of $10^{-5}\%$ of the the radius of the disc.
Similar situation is obtained in three dimensions where a ball $B^3$
is excised. Now the number of negative energy states is $\propto R^2$.

The second example is obtained by the motivation to find out the fate 
of bound states when the disc is squeezed. For this purpose we consider the 
case of excising an elliptical disc from ${\mathbb R}^2$. 
Separation of variables for the Laplacian is possible 
if one uses elliptical coordinates which are given in terms of 
the Cartesian ones by
\beq
x = a\, \cosh \rho \, \cos \theta , \hskip .3in
y = a\, \sinh \rho \, \sin \theta
\label{b4}
\eeq
Here $\rho = \rho_0$ corresponds to the elliptical boundary. In these
coordinates constant $\rho$  curves are ellipses and constant $\theta$
corresponds to hyperbolae orthogonal to the ellipses. Hence 
$\rho_0 \leq \rho < \infty$ and $0 \leq \theta < 2\pi$. The boundary 
condition is 
\beq
\left(\partial_\rho \psi + \kappa \psi\right) \lvert_{\rho_0} = 0
\label{b5}
\eeq
Interestingly the number of bound states decreases as we squeeze
the circular disc and becomes $\propto a\sinh\rho_0$, the minor axis \cite{trgpartharakesh}. 
It is in fact possible to remove all the bound states by squeezing 
sufficiently to lengths $\le \frac{1}{\kappa}$. 
Again the bound states are localised near the boundary.

\section{Bose-Einstein condensation}

The negative energy bound states in the previous section can create 
instabilities when a gas of particles at low temperatures is considered
in such a manifold. The situation is similar to Bose-Einstein condensate
where a divergence in partition function and entropy are prevented 
by a finite number of particles condensing at low temperatures. 
To bring out this comparison, we begin with a brief discussion 
of Bose-Einstein condensation.
Although this is standard textbook material, we want to focus 
attention on some points which can shed light on the problem at hand.

Consider a nonrelativistic gas of bosons, with energy given by
$E_k = k^2 /2 m$. The partition function is given by
$Z = \Tr e^{-\beta (H - \mu N )}$. Normally, we take the states to be of the form
\beq
\vert n_0 , n_{k_1} , n_{k_2}, \cdots \ra 
= { (a^\dagger_0)^{n_0} \over \sqrt{n_0!}}\, { (a^\dagger_{k_1})^{n_{k_1}} \over \sqrt{n_{k_1}!}}
\,{ (a^\dagger_{k_2})^{n_{k_2}} \over \sqrt{n_{k_2}!}}\, \cdots\, \vert 0\ra
\label{a1}
\eeq
Writing $z = e^{\beta \mu}$ for the fugacity, we find
\beq
Z = \prod_{k\neq 0} {1\over (1 - z\,e^ {\beta E_k } )}~ {1\over (1-z)}
\label{a2}
\eeq
The fugacity is in the range $0 \leq z \leq 1$. Notice that there is a singularity in the partition function (and a corresponding logarithimis singularity in the free energy) as
$z \rightarrow 1$. 
The entropy may be evaluated as
\beq
S = {5 \over 2}  {V \over \lambda^3} \, g_{5\over 2} (z) 
- \log (1-z), \hskip .3in g_{\nu } (z) \equiv \sum_1^\infty {z^m \over m^\nu}
\label{a3}
\eeq
Here $\lambda = \sqrt{2 \pi /m T}$ is the thermal wavelength and $V$ is the volume of the system.
There is a singularity in the entropy as well, as $z \rightarrow 1$.
This singularity and the
divergence of the partition function as $z \rightarrow 1$ is taken as the signal for a
phase transition.  To understand the nature of this transition, we restrict
the total number of particles to be $N$. 
It is given in terms of the average occupation numbers as
\beq
N =  \sum_k {z \over  e^{\beta E_k} - z} + {z \over 1-z}
= \sum_k {z \over  e^{\beta E_k} - z} + {\bar n}_0
\label{a4}
\eeq
where ${\bar n}_0 = z/(1-z)$ is the average occupation number in the lowest
eigenstate of the single-particle Hamiltonian, namely, $k =0$.
Working out the integral over $k$, this equation becomes
\beq
1 = {V \over N\, \lambda^3} g_{3\over 2} (z) + {{\bar n}_0 \over N},
\label{a5}
\eeq
As we lower the temperature, the thermal wavelength $\lambda$
increases, lowering the first term on the righthand side, namely, the
contribution of the nonzero modes to this equation.
This can be compensated to some extent by an increase of $z$, which also increases 
$g_{3\over 2} (z)$. However, the maximum value of 
$g_{3\over 2} (z)$ is at $z =1$, $g_{3\over 2} (z) \leq g_{3\over 2} (1) \approx 2.612$.
We see that, at temperatures lower than what is given by this condition,
the first term on the right hand side of
(\ref{a5}) is less than $1$ and the only way to satisfy 
(\ref{a5}) is then for ${\bar n}_0/N$ to be nonzero to make up the deficit.
Thus even in the thermodynamic limit of $N \rightarrow \infty$, there is a nonzero
fraction (in other words a macroscopically significant number)
which must condense into the ground state. The signal for this transition is the singularity in the partition function $Z$ as $z \rightarrow 1$.

The new phase is determined by giving an expectation value to $a_0$, corresponding to the lowest energy sigenstate ($E = 0$).
In other words, rather than states of the form
(\ref{a1}), we take them to be of the form
\beqar
\vert \alpha , n_{k_1} , n_{k_2}, \cdots \ra 
&=&   { (a^\dagger_{k_1})^{n_{k_1}} \over \sqrt{n_{k_1}!}}
\,{ (a^\dagger_{k_2})^{n_{k_2}} \over \sqrt{n_{k_2}!}}\, \cdots\, \vert \alpha, 0\ra
\nonumber\\
a_0 \, \vert \alpha , n_{k_1} , n_{k_2}, \cdots \ra 
&=& \alpha\, \vert \alpha , n_{k_1} , n_{k_2}, \cdots \ra 
\label{a6}
\eeqar
Now the partition function and the entropy become
\beqar
\log Z &=& \alpha^* \alpha \, \log z + {V \over \lambda^3} \, g_{5\over 2} (z)\nonumber\\
S&=& {5 \over 2}  {V \over \lambda^3} \, g_{5\over 2} (z)  + \alpha^* \alpha \, \log z
\label{a7}
\eeqar
The equation for the total number of particles is
\beq
1 =  {V \over N\, \lambda^3} g_{3\over 2} (z) + {\alpha^* \alpha \over N}
\label{a8}
\eeq
This last equation determines $\alpha$. We will get
$\alpha \sim \sqrt{N}\sim \sqrt{V}$, in the thermodynamic limit.
 We see that there is no singularity in
$Z$ or $S$.

The field operator for the particles may be taken as
\beq
\psi (x) = {1\over \sqrt{V}} \sum_k a_k \, e^{-i E_k t + i \vk \cdot \vx} ,
\hskip .3in
\psi^\dagger (x) = {1\over \sqrt{V}} \sum_k a_k \, e^{i E_k t  -i \vk \cdot \vx} 
\label{a9}
\eeq
With $a_0 \sim \alpha \sim\sqrt{V}$, we see that we get a nonzero value
$\la \psi \ra$ for the expectation value of $\psi$ in the ground state of the many-particle system.

It is also useful to consider this in terms of the Euclidean functional integral.
Writing
\beq
\psi(x) = \sum_{-\infty}^\infty \, e^{i \omega_n \tau} \, q_n (\vx )
\label{a10}
\eeq
we find
\beq
Z = \prod_{n,k} {1 \over \E_k + i \omega_n}
\label{a11}
\eeq
where $\E_k = E_k -\mu$. The sum over Matsubara frequencies in $\log Z$ is divergent.
We introduce a Pauli-Villars regulator to write
\beq
{\del \over \del \E_k } \log Z =
- \sum_n \left[ {1\over \E_k + i \omega_n } -   {1\over \E_k + M T  i \omega_n } \right]
\label{a12}
\eeq
This is easily evaluated and leads to
\beq
\log Z = - \sum_k \left[ \log (1- e^{-\beta \E_k})  -  \log (1- e^{-\beta ( \E_k +M T)}) \right]
\label{a13}
\eeq
When the regulator mass $M$ is taken very large, this reduces to the expression
corresponding to $Z$ in (\ref{a2}); thus we mat start from the Euclidean functional integral,
obtain (\ref{a2}) and then carry out the rest of the analysis as done above.
The main point is that the signal for the transition is seen as a singularity of the Euclidean functional integral.
The solution is also given by choosing conditions such that the Euclidean functional integral is well-defined.

Consider now the case where the Laplacian can have negative eigenvalues. It is sufficient to consider
just one such mode to illustrate what happens.
We denote the corresponding energy eigenvalue as $E = - \lambda_1$.
The partition function $Z$ is then given by
\beq
\log Z = - \sum_{k\neq 0} \log (1- z e^{-\beta E_k})
- \log (1-z) - \log (1- z e^{\beta \lambda_1})
\label{a14}
\eeq
We see that we get a singularity even before we get to $z =1$, namely, at
$z \, e^{\beta \lambda_1 } = 1$. Once again, we can take this a signaling a phase transition.
In fact, taking the states to be of the form
\beqar
\vert \alpha_1, n_0 , n_{k_1} , n_{k_2}, \cdots \ra 
&=&  { (a^\dagger_{0})^{n_{0}} \over \sqrt{n_{0}!}}
  { (a^\dagger_{k_1})^{n_{k_1}} \over \sqrt{n_{k_1}!}}
\,{ (a^\dagger_{k_2})^{n_{k_2}} \over \sqrt{n_{k_2}!}}\, \cdots\, \vert \alpha_1, 0\ra
\nonumber\\
a_{\lambda_1} \, \vert\alpha_1, n_0 , n_{k_1} , n_{k_2}, \cdots \ra 
&=& \alpha_1\, \vert \alpha_1, n_0 , n_{k_1} , n_{k_2}, \cdots \ra 
\label{a15}
\eeqar
we find 
\beqar
\log Z &=& \alpha_1^* \alpha_1 \log (z e^{\beta \lambda})  -
\log (1-z)
+ {V \over \lambda^3} \, g_{5\over 2} (z)\nonumber\\
1 &=&  {V \over N\, \lambda^3} g_{3\over 2} (z) + {z \over (1-z )}
+ {\alpha_1^* \alpha_1 \over N}
\label{a16}
\eeqar
The singularity is removed by going to the new phase. We must also consider the value of the fugacity to be in the range $0\leq z \leq z_1$, $z_1 e^{\beta \lambda_1} =1$.

Notice that the conservation of particle number is crucial for this. The value of
$\alpha_1$ have an upper bound by virtue of
(\ref{a16}).
Without such a constraint, or some such constraint arising from a conserved quantum number
(and a corresponding fugacity),
we can have an arbitrary number of particles going into the negative energy state
and creating a theory with no ground state. 
This would be the case, for example, for a relativistic massless scalar field.
Further, in the relativistic case,
the Laplacian occurs under a square root in the expression for the energy.
So negative eigenvalues indicate imaginary energies, rather than negative energies.
The analysis in such cases will have similarities to the present one, but there will also be differences.
We now turn to this problem.

\section{Real scalar field}

We will start by considering a free scalar field theory for which the
 the equation of motion, in Euclidean spacetime, is
given by 
\beq
\square \, \phi = {\del^2 \phi \over \del \tau^2} + \nabla^2\,  \phi  = 0
\label{3}
\eeq
Since boundary considerations are important, the first question is to ask what the action is for which this is the equation of motion.
This is easily seen to be
\beq
\S = {1\over 2} \int (\del \phi )^2 - \oint \del_n \phi \, \phi - {1\over 2} \oint \phi \, \K \, \phi 
\label{4}
\eeq
The variation of this action gives
\beqar
\delta \S &=& \int \delta \phi \, (- \square \phi ) - \oint ( \del_n \delta \phi + \K \delta \phi )\, \phi\nonumber\\
&=&\int \delta \phi \, (- \square \phi ) 
\label{5}
\eeqar
where we have used the self-adjointness of $\K$ and we also take fields and their variations to satisfy the condition
(\ref{2}). This shows that $\S$ is indeed the correct action for the variational derivation of
the equations of motion (\ref{3}).

We can expand the field $\phi$ as
\beq
\phi (x) = \sum_A q_A(\tau )\, u_A(\vx )
\label{6}
\eeq
where $q_A$ can depend on the imaginary time $\tau$ and the modes $u_A(\vx )$ are eigenfunctions of the 
spatial Laplacian,
\beq
- \nabla^2 \, u_A (\vx ) = \omega_A^2 \, u_A (\vx )
\label{7}
\eeq
The use of the mode expansion (\ref{6}) reduces the cation to
\beq
\S = {1\over 2} \sum_A \, \left( {\dot q}_A^2 + \omega_A^2 \, q_A^2 \right)
\label{8}
\eeq
All boundary terms cancel out in the simplification of this expression.

The key point for our analysis is that the
eigenvalues $\omega_A^2$ can be positive or negative. We separate them out as
$\{ u_A \} = ( \{ u_\alpha \}, \{ u_a\} )$, with
the first set corresponding to positive eigenvalues and the second set to negative eigenvalues,
\beqar
-\nabla^2 \, u_\alpha &=& \omega_\alpha^2 \, u_\alpha\nonumber\\
-\nabla^2 \, u_a &=& -\lambda_a^2 \, u_a
\label{9}
\eeqar
with $\omega_\alpha^2$ and $\lambda_a^2$ positive.
The action $\S$ now becomes
\beq
\S =  {1\over 2} \sum_\alpha \, \left( {\dot q}_\alpha^2 + \omega_\alpha^2 \, q_\alpha^2 \right)
+  {1\over 2} \sum_a \, \left( {\dot q}_a^2 - \lambda_a^2 \, q_a^2 \right)
\label{10}
\eeq
The instability is manifest in the last term; the integration of $e^{-\S}$ over the variables
$q_a$ can fail to converge.
As mentioned earlier, we will take the standpoint that the theory must be defined by making the Euclidean functional integral well-defined.
For this, consider periodic boundary conditions in time $\tau$ with period
$\beta = T^{-1}$, $T$ being the temperature. (We use units where the Boltzmann constant $k$ is set to $1$.) Explicitly, we write
\beqar
q_\alpha (\tau ) &=& {1\over \sqrt{\beta}} \sum_n \, q_{\alpha\, n} \, e^{i \Omega_n \tau}
\nonumber\\
q_a (\tau ) &=& {1\over \sqrt{\beta}} \sum_n \, q_{a\, n} \, e^{i \Omega_n \tau}
\label{11}
\eeqar
where $\Omega_n = 2 \pi n \,T$.
Upon using this in (\ref{10}), we see that the first term of the action, namely
$\S_1$, encounters no difficulties. 
The second term $\S_2$ becomes
\beq
\S_2 = \sum_a \sum_{1}^\infty \, q^*_{a\, n} q_{a\, n} \, (\Omega_n^2 - \lambda_\a^2) +
 {1\over 2} \sum_a  q_{a\,0} q_{a\, 0} (- \lambda_a^2)
\label{12}
\eeq
We see that we have stability if we make the restrictions that
$q_{a\, 0} =0$ and that
$\Omega_1^2 \geq \Lambda^2$, where $- \Lambda^2$ is the lowest of the negative eigenvalues.
The last condition means that we have stability only if
\beq
T \geq { \Lambda \over 2 \pi}
\label{13}
\eeq
With these conditions, we can have a well-defined functional integral, the action being given by
\beq
\S = {1\over 2} \sum_\alpha \sum_{0}^\infty \, 
q^*_{\alpha\, n} q_{\alpha\, n} \, (\Omega_n^2 +\omega_\alpha^2 ) +
+ {1\over 2} \sum_a \sum_{1}^\infty \, 
q^*_{a\, n} q_{a\, n} \, (\omega_n^2 - \lambda_\a^2) 
\label{14}
\eeq
The functional integration is convergent. However, it is not enough to
ensure that the partition function is convergent to avoid pathologies.
 We have to make sure the propagators are also well behaved.
We will consider
the calculation of propagators and other correlators to see how a well-defined theory can be
obtained.
The limit of $T \rightarrow \Lambda/ 2\pi$ can then be examined to see
if there is any phase change.

The propagator for the modes of positive eigenvalues is straightforward and gives
\beqar
\la q_\alpha (\tau ) q_\beta(\tau' ) \ra &=&
\delta_{\alpha\beta} { 1\over 2 \omega_\alpha}
\left[ e^{- \omega_\alpha (\tau - \tau' )} \theta (\tau - \tau') + e^{\omega_\alpha (\tau - \tau')}
\theta (\tau' - \tau)\right. \nonumber\\
&&\hskip 1.2in \left. + N_\omega \, \left( e^{- \omega_\alpha (\tau - \tau' )}
+e^{ \omega_\alpha (\tau - \tau' )}\right) \right]
\label{15}
\eeqar
This can be continued to Minkowski signature using $\tau - \tau' \rightarrow i (t-t') $ to get the
corresponding correlator in Minkowski space as
\beqar
\la q_\alpha (t) q_\beta(t' ) \ra &=&
\delta_{\alpha\beta} { 1\over 2 \omega_\alpha}
\left[ e^{- i \omega_\alpha (t-t' )} \theta (t-t' ) + e^{i \omega_\alpha (t -t')}
\theta (t' - t)\right. \nonumber\\
&&\hskip 1.2in \left.+ N_\omega \, \left( e^{- i\omega_\alpha (t- t')}
+e^{i \omega_\alpha (t - t' )}\right) \right]
\label{16}
\eeqar
In (\ref{15}) and (\ref{16}), 
\beq
N_\omega = {1\over e^{\beta \omega_\alpha} - 1}
\label{17}
\eeq
These equations are standard, essentially textbook material.
We now turn to the negative eigenvalues for which we need to evaluate
\beq
\la q_a (\tau ) q^*_b (\tau') \ra
= \delta_{ab}\, {1\over \beta} \,\sum_1^\infty e^{i \Omega_n (\tau - \tau')}{1 \over \Omega_n^2- \lambda_a^2}
\label{18}
\eeq
Recall that the sum does not include the $n= 0$ mode.
This expression can be converted to a contour integral, for $\tau - \tau' > 0$,  as
\beq
\la q_a (\tau ) q^*_b (\tau') \ra
= \delta_{ab} ~{1\over 2}
\left[ \oint_C {dz \over 2\pi} e^{i z (\tau - \tau')} {1\over (z^2 - \lambda_a^2 ) ( e^{i \beta z} -1)}
+ {1\over \beta \lambda_a^2} \right]
\label{19}
\eeq
where the contour $C$ must enclose all the poles of $(e^{i \beta z} - 1)^{-1}$ but
not those which arise from $(z^2 - \lambda_a^2)^{-1}$. 
\begin{figure}[t!]
\begin{center}
\scalebox{.7}{\includegraphics{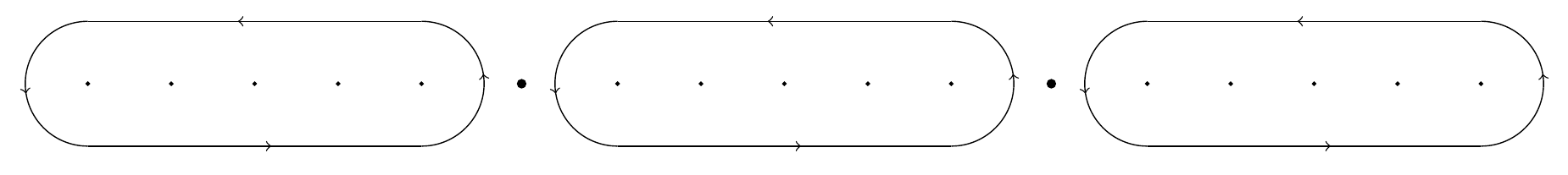}}
\caption{The contour $C$ for the integral in (\ref{19})}
\label{pic1}
\end{center}
\end{figure}
\begin{figure}[b!]
\begin{center}
\scalebox{.75}{\includegraphics{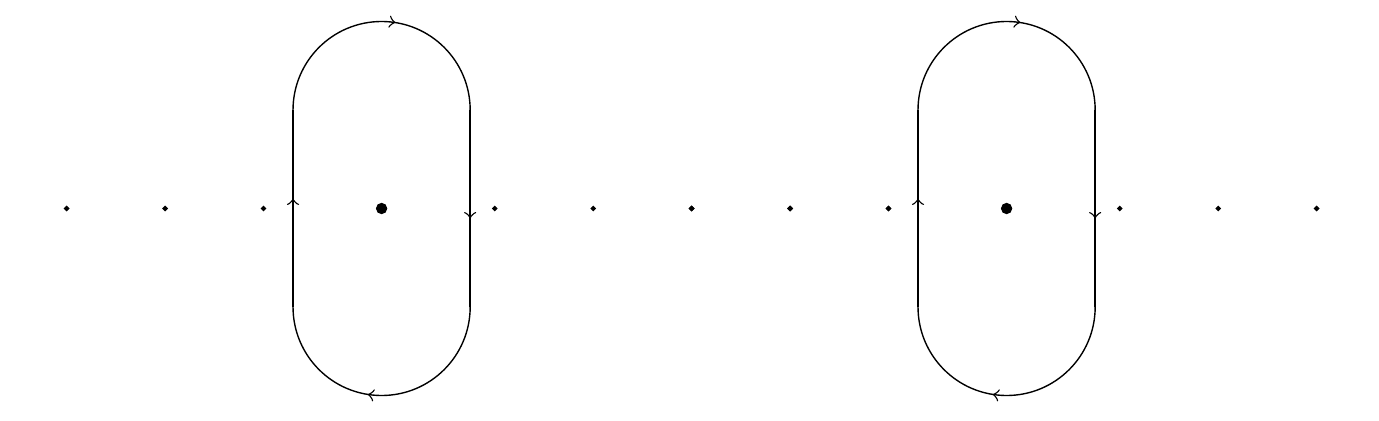}}
\caption{The contour $C_1$ for the integral which gives (\ref{20})}
\label{pic2}
\end{center}
\end{figure}

Unlike the case for
 the positive eigenvalues, we now have additional poles on the real axis due to
 $(z^2 - \lambda_a^2)^{-1}$. So we choose the contour as shown in Fig.\,\ref{pic1}.
  The bold dots are the poles at $z = \pm \lambda_a$, which must be outside the contour.
 The dots at $n = 0, \pm1, \cdots$, are the poles due to $(e^{i \beta z} - 1)^{-1}$.  We can now extend the contours as much as we like into the imaginary directions, since there are no further poles to worry about.
 Further, the factor $e^{i z(\tau - \tau')} \times (e^{i \beta z} - 1)^{-1}$ assures that the integrand falls off exponentially along the imaginary axis.
 Therefore, we can replace the contour $C$ by the new one $C_1$ as shown in Fig.\,\ref{pic2}..
 The contribution is now from the poles at $z = \pm \lambda_a$. We then get
 \beq
\la q_a (\tau ) q^*_b (\tau') \ra
= \delta_{ab} \left\{
{i \over 4 \lambda_a} \left[ { e^{-i \lambda_a (\tau - \tau') }\over e^{-i \beta \lambda_a} - 1}
- { e^{i \lambda_a (\tau - \tau') }\over e^{i \beta \lambda_a} - 1}\right] + {1\over 2 \beta \lambda_a^2}
\right\}, \hskip .3in \tau - \tau' > 0
\label{20}
\eeq
For $\tau - \tau' < 0$, we do not obtain the needed fall-off along the imaginary directions
using $(e^{i \beta z} - 1)^{-1}$. Instead we can use $ e^{i \beta z}\times(e^{i \beta z} - 1)^{-1}$
which has the same poles and residues as $(e^{i \beta z} - 1)^{-1}$.  The rest of the analysis is similar to the case of $\tau - \tau' > 0$ and we get
 \beq
\la q_a (\tau ) q^*_b (\tau') \ra
= \delta_{ab} \left\{
-{i \over 4 \lambda_a} \left[ { e^{-i \lambda_a (\tau - \tau') }\over e^{i \beta \lambda_a} - 1}
- { e^{i \lambda_a (\tau - \tau') }\over e^{- i \beta \lambda_a} - 1}\right] + {1\over 2 \beta \lambda_a^2}
\right\}, \hskip .3in \tau - \tau' < 0
\label{21}
\eeq
The two cases (\ref{20}) and (\ref{21}) can be combined as
\beqar
\la q_a (\tau ) q^*_b (\tau') \ra
&=& \delta_{ab} \Biggl\{ {1\over 2 \beta \lambda_a^2} -
{1\over 8\, \lambda_a\, \sin(\beta \lambda_a /2)}\Bigl[
\exp( - i \lambda_a ( \tau - \tau' ) \pm i \beta \lambda_a /2) \nonumber\\
&&\hskip 1.2in 
+ \exp(  i \lambda_a ( \tau - \tau' ) \mp i \beta \lambda_a /2)\Bigr]\Biggr\}
\label{22}
\eeqar
where the upper sign applies to $\tau - \tau' >0$ and the lower to
$\tau - \tau' < 0$. We may rewrite this also as 
\beq
\la q_a (\tau ) q^*_b (\tau') \ra
= \delta_{ab} \Biggl\{ {1\over 2 \beta \lambda_a^2} -
{1\over 8\, \lambda_a\, \sin(\beta \lambda_a /2)}
\Bigl[
\int dp_0~ e^{ -i p_0 (\Delta - \beta /2)} 
\, \Bigl(\delta(p_0 - \lambda_a ) + \delta (p_0 + \lambda_a)\Bigr)
\Bigr]
\Biggr\}
\label{23}
\eeq
where $\Delta = \tau - \tau'$ and we have only written the case for
$\tau - \tau' > 0$. We can continue this to Minkowski signature by 
the replacements $p_0 \rightarrow i p_0$, $\Delta \rightarrow i (t -t')$.
This leads to the Minkowski space expression
\beq
\la q_a (t ) q^*_b (t') \ra
= \delta_{ab} \Biggl\{ {1\over 2 \beta \lambda_a^2} -
{i\over 4\, \sin(\beta \lambda_a /2)}
\Bigl[
\int dp_0 ~e^{ i p_0 (t -t') \mp p_0 \beta /2} 
\,\, \delta(p_0^2 + \lambda_a^2 ) 
\Bigr]
\Biggr\}
\label{24}
\eeq
We see that there is an exponentially growing part to this and hence there is an instability in processes if we couple this to external sources and consider, for example, a scattering
problem. This can be avoided if we make the following additional rule:
\begin{quotation}
\noindent Observer has access only to the modes $q_{\alpha \, n}$, corresponding to the positive eigenvalues.
\end{quotation}
Thus in any Feynman diagram, we cannot have $q_{a \,n}$ in the external lines or coupling to sources.

We might also worry about possible singularities because of the $\sin (\beta \lambda_a /2)$
in the denominator in (\ref{24}). This can happen for
$T = \lambda_a /2 \pi n$. All such values are excluded already by
(\ref{13}), except for $n =1$ and $\lambda_a = \Lambda$. This last point is the limit of the inequality in (\ref{13}). It is also excluded if we postulate an unattainability rule
that the inequality in (\ref{13}) cannot be saturated, something like a new third law of thermodynamics.
Our conclusion is that the Euclidean functional is well-defined and
the Minkowski continuation of correlators can be meaningfully interpreted if we make the restrictions:
\begin{enumerate}
\item $q_{a\, 0} =0$
\item $T > {\Lambda \over 2 \pi}$, with the limit
$T = {\Lambda \over 2 \pi}$ unattainable
\item Observers have access only to the modes $q_\alpha$, not to $q_a$. However,
$q_a$ can contribute to processes via loops.
\end{enumerate}

We now return to thermodynamic considerations, calculating the free energy and the entropy due to the unstable modes. For the contribution to the free energy, we may write
\beq
\beta \, F = \sum_a \sum_1^\infty \, \log ( \Omega_n^2 - \lambda_a^2)
\label{25}
\eeq
Differentiating with respect to $\lambda_a$ we get a sum similar to what was obtained for the propagators. Carrying out the summation with the same contour integration techniques,
and integrating over $\lambda_a$, we find
\beq
\beta \, F = \sum_a \left[ - \log \lambda_a + \log \left( 2 i \sin (\beta \lambda_a /2)\right)
+ ( \lambda_a {\rm -independent ~term})\right]
\label{26}
\eeq
The constant term can be identified by looking at small values of $\beta \lambda_a$.
This leads to
\beq
F = {1\over \beta} \sum_a \,\log \left[ {\sin (\beta \lambda_a /2) \over (\beta \lambda_a /2)}\right]
\label{27}
\eeq
The entropy can be calculated as
\beq
S = 1 + \sum_a \,\log \left[ {\sin (\beta \lambda_a /2) \over (\beta \lambda_a /2)}\right]
- \sum_a {\beta \lambda_a \over 2} \cot (\beta \lambda_a /2)
\label{28}
\eeq
In both the free energy and the entropy, there is a singularity as
$\beta \lambda_a \rightarrow 2\pi$. For $\beta \lambda_a = 2 (\pi - \epsilon )$,
\beqar
F &\rightarrow& {1\over \beta} \log (\epsilon /\pi)\nonumber\\
S &\rightarrow& {\pi \over \epsilon} + \log \epsilon
\label{29}
\eeqar
We take this as signaling a phase transition. We could consider
 the field as developing an expectation value. However, unlike the case discussed
 in section 2, we do not have a conservation law for the particle number and hence there
 is no equation which can serve to determine the expectation value.
 This is the same problem as in the Bose-Einstein condensation of a free relativistic massless scalar field; the action is of the form $\S = {\half} \int (\del \phi )^2$and the theory can exist
 in a phase with $\la \phi \ra = $ any constant value. If there are interactions, such as a $\phi^4$-term, then the interaction will eventually serve to determine $\la \phi \ra$. This is also the case for a theory with a negative (mass)$^2$ term, which is closer to the situation we have.
 Mass corrections can generically boost the negative eigenvalues to positive or zero values.
 In the context of the $\delta$-function potentials mentioned in the introduction,
 such a mechanism has been studied in \cite{M-CGM} and also in
 \cite{GM-C}. (See also the added reference \cite{ohya}.)
 So to analyze this possibility, we will now consider possible mass corrections arising from a 
 $\phi^4$-interaction.
 
 \section{The interacting theory}
 
 The action for the interacting theory will be taken to be
 \beq
 \S = \sum_\alpha \sum_1^\infty \, q^*_{\alpha\, n} q_{\alpha \, n} \, (\Omega_n^2 + \omega_\alpha^2)
 + {1\over 2} \sum_\alpha q_{\alpha \, 0}^2 \, \omega_\alpha^2
 +  \sum_a \sum_1^\infty \, q^*_{a\, n} q_{a \, n} \, (\Omega_n^2 - \lambda_a^2)
 + g \int \phi^4
 \label{30}
 \eeq
 We will separate out the unstable modes by writing 
 $\phi = \vf + \eta$, where
 \beq
 \vf = {1\over \sqrt{\beta}} \sum q_{\alpha \, n} e^{i \Omega_n \tau} \, u_\alpha (\vx ),
 \hskip .3in
 \eta = {1\over \sqrt{\beta}} \sum q_{a\, n} e^{i \Omega_n \tau} \, u_a (\vx )
 \label{31}
 \eeq 
The strategy is to integrate out the $\eta$'s to obtain an effective action for the
$\vf$'s. This result can then be continued to Minkowski space
and real-time processes can be calculated.
So long as there are no sources coupled to the unstable modes,
$\eta$'s only contribute in loops and this process can be consistently implemented.

First of all, let us consider tree-graphs where the $\eta$-propagator can occur.
The question is whether these can lead to new instabilities requiring new restrictions.
Consider as an example the term
$\int \vf(x)^3 \,\la \eta (x) \eta (y) \ra \, \vf(y)^3$. Using the expression for the propagator
in (\ref{23}), this can be evaluated in a straightforward manner as
\beq
\int \vf(x)^3 \,\la \eta (x) \eta (y) \ra \, \vf(y)^3
= \int d^4x d^4y~\vf (x)^3 \, V(x, y)\, \vf (y)^3\label{32}
\eeq
where
\beq
V(x,y) = \sum_a u_a(\vx) \, u_a^*(\vy ) \, \int {d \omega \over 2 \pi}{d \omega' \over 2 \pi} \,
e^{i \omega x^0 + i \omega' y^0} \left[
{(2\pi)^2 \delta (\omega) \delta (\omega') \over 2 \beta \lambda_a^2}
- {2 \pi \delta (\omega + \omega') \over 2 (\omega^2 + \lambda_a^2)}
\right]
\label{33}
\eeq
There is nothing pathological about this. Notice that the first term in
$V(x,y)$ is an instantaneous potential which is also temperature-dependent.
It is confined to a region close to the boundary since the $u_a(\vx)$ fall off as we move away from the boundary.
Turning to loop corrections, the
simplest one we can evaluate is the one-loop mass correction due to the unstable modes.
This is easily seen to be given by
\beqar
\Delta \S &=& {1\over 2} \int  V(x)\, \vf^2 (x)\nonumber\\
V(x) &=& 12 g \sum_a u_a(\vx) \, u^*_a (\vx) \left[
{1\over 2 \beta \lambda_a^2} -{1\over 4 \lambda_a } \cot (\beta \lambda_a /2)\right]
\label{34}
\eeqar
Being a position-dependent mass term, this is really a single-particle potential for the
$\vf$ modes. $V(x)$ is again concentrated near the boundary.
It is positive for all values of $\beta \lambda_a$ in the range of interest.
Thus $\vf$'s experience a repulsive potential near the boundary helping to avoid any further instabilities, at least to this order.

\begin{figure}[t!]
\begin{center}
\scalebox{1}{\includegraphics{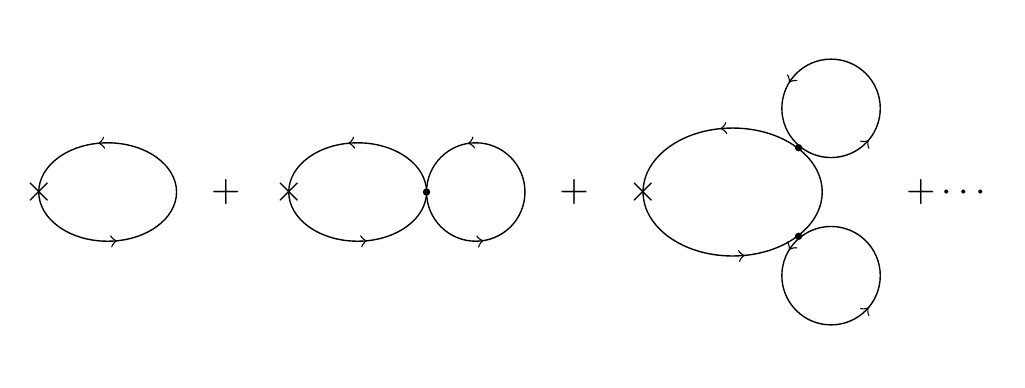}}
\caption{Diagrams involving the propagator of the unstable modes
which contribute to the partition function}
\label{pic3}
\end{center}
\end{figure}

Let us now consider the thermodynamic quantities. The partition function will get
contributions from diagrams of the type shown in Fig.\,\ref{pic3}, where the propagators are those corresponding to the unstable modes $\eta$. The first term is the free part which gives the expressions (\ref{27}) and (\ref{28}).
The extra loops correspond to the modification of the propagator via a mass correction for the
$\eta$ fields. So while this does not have to be taken account of in
external lines, this mass correction does influence the thermodynamics.
Evidently, the mass correction for this is of the form ${\half} \int  V(x) \,\eta^2$.
Thus the effect of the series is to change the expression for the free energy to
\beq
\beta \, F = \sum_a \sum_1^\infty \log (\Omega_n^2 - {\tilde \lambda}_a^2 )
\label{35}
\eeq
where $\tilde\lambda_a$ are the eigenvalues
of $-\nabla^2 + V(x)$,
\beq
(- \nabla^2 + V(x) ) \, {\tilde u}_a =
-{\tilde \lambda}_a^2 \, {\tilde u}_a
\label{36}
\eeq
with $V(x)$ as given in (\ref{34}). This extra repulsive potential can make the eigenvalues
$-{\tilde \lambda}_a^2$ positive avoiding the singularity  as $\beta \lambda_a \rightarrow 2 \pi$.
Unfortunately, an explicit calculation is rather difficult,
since the potential $V(x)$ diverges as $\beta \lambda_a \rightarrow 2 \pi$, making any perturbative evaluation of corrections inadequate. Nevertheless, it is useful to get an estimate of the correction to the eigenvalues in perturbation theory, say, to first order.
For the example introduced in Sec. 2 of ${\mathbb R}^2$ with a disc of radius $R$ excised from it,
consider the case
when we have only one eigenstate with negative eigenvalue.
This can be obtained for $\kappa R = 0.5$ for example. Taking this as an illustrative case,
we find $z_* \approx 0.165725$ and
the eigenfunction is
\beqar
u (x) &=& (2 \pi R^2 {\cal I})^{-1/2} \, K_0 (z_* r /R)\nonumber\\
{\cal I}&=& \int_1^\infty dw\, w\, (K_0 (z_* w) )^2 
\label{36a}
\eeqar
The eigenvalue $-z_*^2 /R^2$, with the first correction included becomes
\beq
- {\tilde \lambda}^2 \approx
{z_*^2 \over R^2} \left[ - 1 + 
(0.2102)\times {3 \, g\over \lambda} \left[ {1\over (\beta \lambda /2)} - \cot (\beta \lambda /2)\right]
\right]
\label{36b}
\eeq
The correction is not small except for very small $g$ and $\beta \lambda$, and 
diverges as $\beta \lambda \rightarrow 2 \pi$, showing that the negative modes are eliminated
 before we get to $\beta \lambda = 2 \pi$.
 As mentioned above, perturbation theory is not adequate for this analysis; we plan to explore
 a numerical approach to this question in future.
 
\section{Discussion}

In this paper we have considered field theories on manifolds with boundaries and for which the one-particle kinetic energy operator, taken as the Laplacian, has negative eigenvalues.
The possibility of negative eigenvalues is related to the choice of boundary conditions.
The general theory shows that there is a large class of boundary conditions, in fact infinitesimally close to standard and well-known ones such as Dirichlet and Neumann, for which the Laplacian
can have negative eigenvalues.
Explicit illustrative examples were given in Sec. 2.
Our strategy for analyzing such theories was to start with defining the theory at finite and high enough temperature for which we have a well-defined partition function and then pose the
question of whether we can lower the temperature to zero.
The analysis leads to three different cases.

If we consider free field theories with a conserved particle number operator,
there is Bose-Einstein condensation as we lower the temperature with a thermodynamically nontrivial fraction condensing into the mode with the lowest
(and negative) eigenvalue for the Laplacian; the transition temperature is related to this lowest eigenvalue.
Stability in this case is due to the total number of particles being fixed.

By contrast, if we consider a real scalar field, for which there is no conserved number operator
(and hence no corresponding chemical potential), we find that a stable theory is possible
only if the temperature remains above a certain value $\Lambda/ 2 \pi$, where $- \Lambda^2$ is the lowest eigenvalue of the Laplacian.
The temperature $\Lambda / 2 \pi$ plays the role of absolute zero for this case, with a corresponding unattainability condition (as in the usual third law of thermodynamics).
Further, observers should have access only to the modes with positive eigenvalues.
These features, particularly the fact that the system shows finite temperature and that the modes of negative eigenvalues are localized near the boundary, are very suggestive of what is observed in the region outside the horizon of a black hole.
At this point, this is still an intriguing analogy; the possibility of a deeper connection
is worth exploring. As the temperature approaches the value $\Lambda/ 2\pi$, thermodynamic quantities such as the free energy and entropy diverge again suggesting a phase transition.
However, within the free theory, there is no way to determine the expectation value for the
field.

The third case of interest, which is also related to the second case, is when we have an interacting scalar field theory. We considered a simple example of a $\phi^4$-type interaction.
The modes with negative eigenvalues can contribute in loops and the general effect is to create a new repulsive potential near the boundary. We expect this to eliminate
the negative eigenvalues and lead to a stable theory as the temperature approaches the critical value $\Lambda / 2 \pi$.
The extra potential diverges as this limit is approached, making any perturbative analysis nonviable.
We plan to explore this question is more detail numerically in future work.

The addition of `mass term' on the boundary would presumably save the thermodynamic
quantities such as the free energy. But the negative energy modes localised  at the boundary
remain as `zero energy' modes. The lesson is there could be edge states localised at the boundary even in conventional 
circumstances. But under perturbation through interactions they will go away since there is `no gap' with bulk modes.
But with new boundary conditions which have global origin 
they will remain stable. This will be seen in the behaviour of 
two-point functions of the edge modes.

Lastly we would like to remark about fermionic theories . The Dirac operator on such manifolds have 
to be supplemented by Atiyah, Patodi and Singer global boundary conditions \cite{atiyah} in order to be self adjoint. 
The square of the Dirac operator is positive definite. But edge states localised on the boundary persists, see for example \cite{balrecent}.
This can change thermodynamics,  a question which we plan to
explore in future.   

After this paper was written, we became aware of
\cite{ohya} 
where a specific realization of the negative eigenvalues is used as a possible mechanism for breaking gauge symmetries.
(We thank S. Ohya fro bringing this work to our attention.)
Our analysis is quite different, even though there are some points of overlap.
We consider the question of how the scalar field self-interactions affect the whole issue of 
condensation to be not settled in that case as well.
\vskip .1in
\noindent {\bf Acknowledgements}

\vskip .1in
We thank Professor A.P. Balachandran for discussions.
TRG would like to thank Manuelo Asorey for sharing
his insights.

This research was also supported by the U.S.\ National Science
Foundation grant PHY-1213380
and by a PSC-CUNY award. 



\end{document}